\documentclass[aps,prd,preprint,nofootinbib,superscriptaddress]{revtex4-1}
\pdfoutput=1
\newcount\rem \rem=1
\ifnum \rem=0 
\newcommand{\ygn}[1]{}
\newcommand{\adn}[1]{}
\newcommand{\mgn}[1]{}
\newcommand{\sts}[1]{}
\newcommand{\tk}[1]{}
\else
\newcommand{\ygn}[1]{{\bf \color{red} [YG:~#1]}}
\newcommand{\adn}[1]{{\bf \color{blue} [AD:~#1]}}
\newcommand{\mgn}[1]{{\bf \color{green} [MG:~#1]}}
\newcommand{\sts}[1]{{\bf \color{magenta} [STS:~#1]}}
\newcommand{\tk}[1]{{\bf \color[rgb]{0.14,0.62,0.63} [TK:~#1]}}
\fi

\newcounter{mycount}
\newcommand{\pauseen}{\setcounter{mycount}{\value{enumi}}\end{enumerate}}
\newcommand{\resumeen}{\begin{enumerate}[(i)]\setcounter{enumi}{\value{mycount}}}

\usepackage{amssymb,amsmath}
\usepackage{color}
\usepackage[toc,page]{appendix}
\usepackage{graphicx}
\usepackage{slashed}
\usepackage{subfigure}
\usepackage{hyperref}
\usepackage{enumerate}
\usepackage{url}
\usepackage{ulem}

\newcommand{\nn}{\nonumber}

\newcommand{\ie}{{\em i.e.}}

\newcommand{\ket}[1]{|{#1} \rangle}

\newcommand{\gmu}{\gamma^{\mu}}

\newcommand{\gfive}{\gamma^5}

\def\beq{\begin{eqnarray}}
\def\eeq{\end{eqnarray}}

\begin{document}
\preprint{KEK-TH-2469}

\title{A Precision Relation between \texorpdfstring{\boldmath{$\Gamma(K\to\mu^+\mu^-)(t)$}}{Gamma(K to mu mu)(t)}  and 
\texorpdfstring{\boldmath{${\cal B}(K_L\to\mu^+\mu^-)/{\cal B}(K_L\to\gamma\gamma)$}}{B(KL to mu mu)/B(KL to gamma gamma)}
}

\author{Avital Dery}
\email{avital.dery@cornell.edu}
\affiliation{Department of Physics, LEPP, Cornell University, Ithaca, NY 14853, USA}
\author{Mitrajyoti Ghosh}
\email{mg2338@cornell.edu}
\affiliation{Department of Physics, LEPP, Cornell University, Ithaca, NY 14853, USA}
\author{Yuval Grossman}
\email{yg73@cornell.edu}
\affiliation{Department of Physics, LEPP, Cornell University, Ithaca, NY 14853, USA}
\author{Teppei Kitahara}
\email{teppeik@kmi.nagoya-u.ac.jp}
\affiliation{Institute for Advanced Research, Nagoya University, Nagoya 464–8601, Japan}
\affiliation{Kobayashi-Maskawa Institute for the Origin of Particles and the Universe,
Nagoya University, Nagoya 464–8602, Japan}
\affiliation{KEK Theory Center, IPNS, KEK, Tsukuba 305--0801, Japan}
\affiliation{CAS Key Laboratory of Theoretical Physics, Institute of Theoretical Physics, Chinese Academy of Sciences, Beijing 100190, China}
\author{Stefan Schacht}
\email{stefan.schacht@manchester.ac.uk}
\affiliation{Department of Physics and Astronomy, University of Manchester, Manchester, M13 9PL, United Kingdom\\
\vspace{1.2cm}}

\begin{abstract}
\noindent
We find that the phase appearing in the unitarity relation between $\mathcal{B}(K_L\rightarrow \mu^+\mu^-)$ and $\mathcal{B}(K_L\rightarrow \gamma\gamma)$ is equal to the phase shift in the interference term of the time-dependent $K\rightarrow \mu^+\mu^-$ decay. A probe of this relation at future kaon facilities constitutes a Standard Model test with a theory precision of about $2\%$. 
The phase has further importance for sensitivity studies regarding the measurement of the time-dependent   $K\rightarrow \mu^+\mu^-$ decay rate to extract the CKM matrix element combination $\vert V_{ts} V_{td} \sin(\beta+\beta_s)\vert\approx A^2\lambda^5\bar\eta$. We find a model-independent theoretically clean prediction, $\cos^2\varphi_0 = 0.96 \pm 0.03$. The quoted error is a combination of the  theoretical and experimental errors, and both of them are expected to shrink in the future. Using input from the large-$N_C$ limit within chiral perturbation theory, we find a theory preference towards solutions with negative  $\cos\varphi_0$, reducing a four-fold ambiguity in the angle $\varphi_0$ to a two-fold one. 
\end{abstract}

\maketitle

%%%%%%%%%%%%%%%%%%%
\section{Introduction}\label{sec:intro}
%%%%%%%%%%%%%%%%%%%
%
A recent proposal has shown that short-distance  parameters of the decay $K\to\mu^+\mu^-$ can be cleanly extracted from a measurement of the $K_L\text{--}K_S$ time-dependent rate~\cite{DAmbrosio:2017klp,Dery:2021mct,Brod:2022khx}.
The time-dependent rate for a beam of initial $K^0$ particles can be written as
\begin{eqnarray}\label{eq:timeDep}
    \frac{1}{{\cal N}}\frac{d\Gamma(K^0\to\mu^+\mu^-)}{dt} \,  = f(t) \equiv \, C_L \,e^{-\Gamma_L t} + C_S \,e^{-\Gamma_S t} +  2\,C_\text{Int.}\cos(\Delta M_K t-\varphi_0) e^{-\frac{\Gamma_L+\Gamma_S}{2} t}\, ,
\end{eqnarray}
where ${\cal N}$ is a normalization factor, $\Gamma_L$ ($\Gamma_S$) is the $K_L$ ($K_S$) decay width, and $\Delta M_K$ is the $K_L\text{--}K_S$ mass difference. 
Then, the four experimental parameters characterizing the time dependence,
\begin{equation}
    \left\{C_L, \, C_S, \, C_\text{Int},\, \varphi_0 \right\} \, ,
\end{equation}
are directly related to the four theory parameters describing the system~\cite{Dery:2021mct}, 
\begin{equation}
    \left\{ |A(K_S)_{\ell=0}|,\,\, |A(K_L)_{\ell=0}|, \,\, |A(K_S)_{\ell=1}|, \,\,\arg\big[A(K_S)_{\ell=0}^* \, A(K_L)_{\ell=0}\big]\right\}\, ,
\end{equation}
where the subscripts $\ell=0$ ($s$-wave symmetric wave function) and $\ell=1$ ($p$-wave  anti-symmetric wave function) correspond to the CP-odd and -even $(\mu^+\mu^-)$ final states, respectively. 
The relations between the experimental and theory parameters are given by
\begin{align}\label{eq:match-th-exp}
\begin{aligned}
    C_L\,\,\,\, &= \, |A(K_L)_{\ell=0}|^2\, , \\ 
    C_S\,\,\,\, &= \, |A(K_S)_{\ell=0}|^2 + \beta_\mu^2 |A(K_S)_{\ell=1}|^2 \, , 
    %\qquad \beta_\mu = \left(1-\frac{4m_\mu^2}{m_K^2}\right)^{1/2}\,,
    \\ 
    C_\text{Int.}\, &= \, |A(K_S)_{\ell=0}^* \, A(K_L)_{\ell=0}|  = |A(K_S)_{\ell=0}| |A(K_L)_{\ell=0}| \,, \\
    \varphi_0 \,\,\,\,\, &= \, \arg\big[A(K_S)_{\ell=0}^* \, A(K_L)_{\ell=0}\big]\, ,
\end{aligned}
\end{align}
with
\begin{align}
    \beta_\mu = \sqrt{1-\frac{4m_\mu^2}{m_{K^0}^2}}\,.
\end{align}

The experimental parameter $\varphi_0$, which is the phase shift of the oscillating rate in Eq.~\eqref{eq:timeDep}, is a combination of the relative weak and strong phases 
between the $K_S$ and $K_L$ amplitudes to the CP-odd final state. 
In this paper, we demonstrate that this phase shift is closely related to the proportionality coefficient in the ratio between the rates of $K_L\to\mu^+\mu^-$ and $K_L\to\gamma\gamma$. 

The ratio between the rates of $K_L\to\mu^+\mu^-$ and $K_L\to\gamma\gamma$ is of historical significance. Using CPT invariance, unitarity, and the well-motivated assumption that the absorptive part of the $K_L \to \mu^+ \mu^-$ amplitude is dominated by the two-photon intermediate state, the ratio between the rates of $K_L\to\mu^+\mu^-$ and $K_L\to\gamma\gamma$ is bounded by the lower limit \cite{Christ:1971hr, Stern:1973xs, Sehgal:1966wr, Sehgal:1969zok} 
\beq 
\label{eq:RKL}
R_{K_L} \equiv \frac{\Gamma(K_L \to \mu^+\mu^-)}{\Gamma (K_L \to \gamma \gamma)} 
\geq 1.195 \times 10^{-5} \,. \
\eeq
However, back in the 1970's, this conflicted with the contemporary experimental upper bound of~$0.4 \times 10^{-5}$~\cite{Clark:1971kj,Arnold:1968zza}, leading to the so-called \lq\lq{}$K_L \to \mu^+\mu^-$ puzzle\rq\rq{}~\cite{Stern:1973xs,Chen:1971kfc} that gained much attention. But today, this ratio is measured in the experiment as~\cite{Workman:2022ynf} 
\begin{align}
R^{\mathrm{exp}}_{K_L} = (1.250 \pm 0.022) \times 10^{-5}\,. 
\end{align}
This means that the observed branching ratio ${\cal B} (K_L \to \mu^+ \mu^-) = (6.84 \pm 0.11) \times 10^{-9}$~\cite{Workman:2022ynf}, along with ${\cal B} (K_L \to \gamma \gamma) = (5.47 \pm 0.04) \times 10^{-4}$~\cite{Workman:2022ynf} known today  obey the lower limit prescribed by CPT invariance and unitarity.

In this paper, we show that the phase $\varphi_0$ is cleanly predicted in the Standard Model (SM), up to a four-fold discrete ambiguity, making its measurement a potent test of the SM. The discrete ambiguity can be partially resolved by using further theory input from the literature in the large-$N_C$ limit of chiral perturbation theory (ChPT). This result is additionally significant for sensitivity estimations of a future measurement of the short-distance parameters.   
 
 Leptonic kaon decays have been a field that received a lot of attention in the literature recently. Effects from CPV in kaon mixing on $K\rightarrow \mu^+\mu^-$ have been taken into account in Refs.~\cite{DAmbrosio:2017klp, Brod:2022khx}, and implications for physics beyond the SM have been studied in Refs.~\cite{Chobanova:2017rkj,Endo:2017ums,Dery:2021vql}. 
 Another future high precision test of the SM employing the ratio $\mathcal{B}(K_S\rightarrow \mu^+\mu^-)_{\ell=0}/\mathcal{B}(K_L\rightarrow \pi^0\nu\bar{\nu})$ has been identified in Ref.~\cite{Buras:2021nns}. 
  Advances in calculating $K_L\rightarrow \mu^+\mu^-$ and $K_L\rightarrow \gamma\gamma$ on the lattice can be found in Refs.~\cite{Christ:2020bzb, Zhao:2022pbs, Christ:2022rho}. On the experimental side, the LHCb collaboration recently found an improved bound on $K_S\rightarrow \mu^+\mu^-$~\cite{LHCb:2020ycd} and $K_{S,L}\rightarrow 2(\mu^+\mu^-)$~\cite{Gomez:2022}.
 
  In Sec.~\ref{sec:notation} we introduce our notation and summarize key results from the literature. In Sec.~\ref{sec:model-independent} we determine $\cos^2\varphi_0 $, which predicts $\varphi_0$ up to a four-fold ambiguity, in a model-independent way, only assuming that the long-distance contributions are SM-like. In Sec.~\ref{sec:largeNC} we reduce this ambiguity to a two-fold one by using the large-$N_C$ limit and assuming that the short-distance physics is known.
 We demonstrate that the remaining ambiguity cannot be resolved using current knowledge in Sec.~\ref{sec:determination}. 
 We conclude in Sec.~\ref{sec:conclusion}.

%%%%%%%%%%%%%%%%%%%%%%%%
\section{Setup and notation \label{sec:notation}}
%%%%%%%%%%%%%%%%%%%%%%%%

\begin{figure}[t]
\begin{center}
\includegraphics[width=0.5\textwidth]{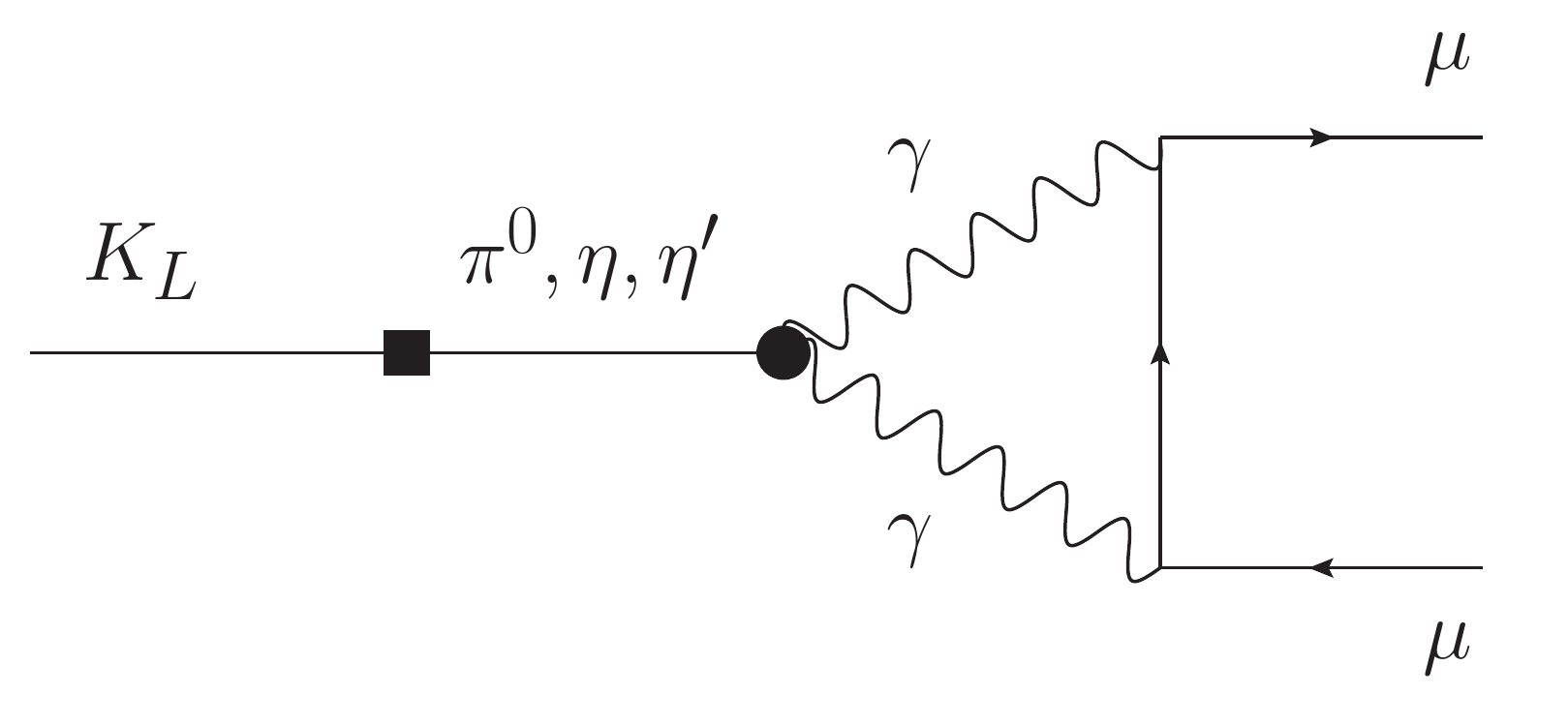}
\caption{Leading order Feynman diagram of the long-distance two-photon contribution to $K_L\rightarrow \mu^+\mu^-$. 
\label{fig:two-photon-contribution}}
\end{center}
\end{figure} 

We work within a framework defined by the following approximations, as detailed in Ref.~\cite{Dery:2021mct}:
\begin{enumerate}[(i)]
\item We neglect CPV in mixing, which is a sub-dominant effect for our purposes. Note that when considering higher order corrections, this effect can be taken into account consistently~\cite{Brod:2022khx}. 
\item We neglect CPV in the long-distance contribution.
\item We assume that the leptonic current is (axial-)vectorial, \ie, given by $\bar{\mu} \gmu (a+b\gfive) \mu$. 
\pauseen
In the SM,  all three approximations are fulfilled  within the precision relevant for our findings. 
In addition, in the following we assume that 
\resumeen
\item
The long-distance contribution to the $K_L\to\mu^+\mu^-$ amplitude is SM-like. That is, the only non-negligible intermediate state is the di-photon state.
\pauseen
Regarding the short-distance contribution, we explicitly state whenever our results are relevant regardless of any assumption on the nature of the short-distance physics, and when SM input is used.

Adopting this setup, it has been shown that the decay of $K_S$ to the CP-odd final state, $(\mu^+\mu^-)_{\ell=0}$ involves only short-distance physics. Additionally, the $K_L\to\mu^+\mu^-$ decay proceeds only to the CP-odd final state, $(\mu^+\mu^-)_{\ell=0}$. However, the latter involves two contributions of different underlying physics:
\begin{enumerate}
\item 
A short-distance (SD) contribution, arising to leading order from box and electroweak penguin diagrams, for which the ingredients for a precise SM prediction are straight-forward to derive.
\item 
A long-distance (LD) contribution, strongly dominated by the on-shell two photon intermediate state, see Fig.~\ref{fig:two-photon-contribution}.   
\end{enumerate}  
The on-shell two photon contribution is known to be significantly larger in magnitude than both the SD contribution and the off-shell part of the LD contribution.

The $K^0\to (\mu^+\mu^-)_{\ell=0}$ and $\overline{K}^0\to (\mu^+\mu^-)_{\ell=0} $ amplitudes can be written as a sum of two general contributions, with corresponding weak and strong phases, 
\begin{eqnarray}
    A_{\ell=0} &=& |A_{SD}| e^{i\theta_{SD}}e^{i\delta_{SD}} + |A_{LD}|e^{i\theta_{LD}}e^{i\delta_{LD}}\, , \\ \nonumber
    \overline A_{\ell=0} &=& -\left(|A_{SD}| e^{-i\theta_{SD}}e^{i\delta_{SD}} + |A_{LD}|e^{-i\theta_{LD}}e^{i\delta_{LD}}\right)\, ,
\end{eqnarray}
where the overall minus sign for $\overline A_{\ell=0}$ is due to the CP nature of the final state.   
Using the convention
\beq
\ket{K_S} = p\ket{K^0} + q\ket{\overline{K}^0} \,, 
\qquad
\ket{K_L} = p \ket{K^0} - q \ket{\overline{K}^0}\,,
\eeq
the mass   eigenstate amplitudes are related to $A_{\ell=0}$ and $\overline{A}_{\ell=0} $ by~\cite{Dery:2021mct}
\begin{eqnarray}
    A(K_S)_{\ell=0} &=& \frac{1}{\sqrt{2}}\left[A_{SD}(1+\lambda_{SD})+A_{LD}(1+\lambda_{LD})\right]\,,\\ \nonumber
    A(K_L)_{\ell=0} &=&  \frac{1}{\sqrt{2}}\left[A_{SD}(1-\lambda_{SD})+A_{LD}(1-\lambda_{LD})\right]\,,
\end{eqnarray}
where
\begin{align}
A_{SD} &\equiv \vert A_{SD}\vert e^{i\theta_{SD}} e^{i\delta_{SD}}\,,  &
A_{LD} &\equiv \vert A_{LD}\vert e^{i\theta_{LD}} e^{i\delta_{LD}}\,, \\
%%%
\overline{A}_{SD} &\equiv -\vert A_{SD}\vert e^{-i\theta_{SD}} e^{i\delta_{SD}}\,,  &
\overline{A}_{LD} &\equiv -\vert A_{LD}\vert e^{-i\theta_{LD}} e^{i\delta_{LD}}\,, 
\end{align}
and
\begin{equation}
    %\lambda_\ell \equiv \frac{q}{p}\frac{\overline A_\ell}{A_\ell}\, , \qquad 
    \lambda_{SD}\equiv  \frac{q}{p}\frac{\overline A_{SD}}{A_{SD}}\, , \qquad \lambda_{LD}\equiv  \frac{q}{p}\frac{\overline A_{LD}}{A_{LD}}\, .
\end{equation}

As we focus here mainly on amplitudes with $\ell=0$ final states, compared to the notation of Ref.~\cite{Dery:2021mct} we use for brevity the notation
\begin{align}
\lambda_{SD} &\equiv \lambda_0^{SD}\,, &
\lambda_{LD} &\equiv \lambda_0^{LD}\,, \\
A(K_S) &\equiv A(K_S)_{\ell=0}\,, &
A(K_L) &\equiv A(K_L)_{\ell=0}\,.
\end{align}
The amplitudes are normalized such that
\begin{align}
\mathcal{B}(K_{S,L}\rightarrow \mu^+\mu^-)_{\ell=0} &= \frac{\tau_K \beta_{\mu}}{16\pi m_K} \vert A(K_{S,L})\vert^2\,.
\end{align}
We also use
\begin{align}
\mathcal{B}(K_{L}\rightarrow \gamma\gamma) &= \frac{\tau_K}{32\pi m_K} \vert A(K_L\rightarrow \gamma\gamma)\vert^2\,.
\label{eq:BKLgg}
\end{align}

Up to this point, the labels LD and SD are just naming. However, in the following we will treat them as corresponding to what we think of as long-distance and short-distance amplitudes. It is important to note that the separation into LD and SD contributions is not well-defined. 
We think of short-distance physics as having no sources for a strong phase, while long-distance physics can go on-shell. However, any on-shell intermediate state can also be considered to contribute off-shell. Therefore there is no way to unambiguously define the separation.
In the following we keep the strong phases general, while we insert knowledge of the SM weak phase of the long-distance contribution.

We then have~\cite{Dery:2021mct} 
\begin{align}
\lambda_{LD} &= -\left(\frac{V_{cd} V_{cs}^*}{V_{cd}^* V_{cs}}\right)\left(\frac{V_{ud}^* V_{us} }{V_{ud} V_{us}^*}\right)
= -e^{-2i \theta_{uc}}\,. \label{eq:lambdaLD}
\end{align}
Equation~(\ref{eq:lambdaLD}) corrects a typo in Eq.~(42) of Ref.~\cite{Dery:2021mct}.
Here, 
\begin{align}
\theta_{uc} \equiv \mathrm{arg}\left(-\frac{V_{cd} V_{cs}^*}{V_{ud} V_{us}^*}\right) = \mathcal{O}(\lambda^4)\,,
\end{align}
such that, to $\mathcal{O}(\lambda^4)$,
\begin{align}\label{eq:weakphases}
    \arg\lambda_{LD} = -2\theta_{LD} +\pi = \pi\,,
\end{align} 
that is
\begin{align}
\theta_{LD} &= 0\,, \qquad
\lambda_{LD} = -1\,.
\end{align}

No assumptions are made for the short-distance phase,
\begin{align}
    \arg\lambda_{SD} =  -2\theta_{SD}+\pi \, .
\end{align}
We have therefore
\begin{align}
A(K_S) &= \frac{1}{\sqrt{2}} A_{SD} (1+\lambda_{SD}) = i \sqrt{2} |A_{SD}|\sin\theta_{SD} e^{i\delta_{SD}}\,, \\
A(K_L) &= \frac{1}{\sqrt{2}} \left[ A_{SD} (  1-\lambda_{SD}) + 2 A_{LD}\right] = \sqrt{2}\Big(|A_{LD}|e^{i\delta_{LD}} + |A_{SD}|\cos\theta_{SD}e^{i\delta_{SD}}\Big) \,.
\label{eq:AKL}
\end{align}
Note that $A(K_S)$ is pure short-distance and is manifestly CP-odd. 
The oscillation term in the rate is then controlled by the interference term:
\begin{eqnarray}\label{eq:AKSstarAKL}
    A(K_S)^*A(K_L) &=& -2i|A_{SD}|\sin\theta_{SD}\left(|A_{LD}|e^{i\Delta\delta}+ |A_{SD}|\cos\theta_{SD}\right) \, ,
\end{eqnarray}
where $\Delta\delta \equiv \delta_{LD}-\delta_{SD}$.

%%%%%%%%%%%%%%%%%%%%%%%%%%%%%%%%%%%%%
\section{Determination of \texorpdfstring{\boldmath{$\cos^2\varphi_0$}}{cos2 phi0}: model-independent  
\label{sec:model-independent}}
%%%%%%%%%%%%%%%%%%%%%%%%%%%%%%%%%%%%%
%
From Eqs.~(\ref{eq:match-th-exp}) and (\ref{eq:AKSstarAKL}), we have  
\begin{align}
\cos^2\varphi_0\, = \, \frac{{\rm Re}\big[A(K_S)^* A(K_L)\big]^2}{|A(K_S)A(K_L)|^2} %\\
    = \frac{(\sqrt{2}|A_{LD}|\sin\Delta\delta)^2}{|A(K_L)|^2}\ %\\
    = \frac{\big[A(K_L)_\text{absorptive}\big]^2}{|A(K_L)|^2} \,, \label{eq:cosGeneral}
\end{align}
where
\begin{equation}
    A(K_L)_\text{absorptive} \equiv  {\rm Im}\big[A(K_L)\big]\, ,
    \label{eq:AKL-absorptive}
\end{equation}
and we define the imaginary part relative to the strong phase of the SD amplitude (or in the basis in which the SD amplitude carries no strong phase, hence $\delta_{LD}=\Delta\delta$).
This convention corresponds to choosing $C_{\text{had.}}$, introduced in Sec.~\ref{sec:largeNC} below, to be real without loss of generality.

The numerator is simply the on-shell long-distance contribution to the $K_L$ amplitude.
Under the assumption that the only non-negligible intermediate state is the di-photon state~\cite{Martin:1970ai}, this absorptive part is equal to the discontinuity of the three-point diagram (see Fig.~\ref{fig:two-photon-contribution}), 
\begin{eqnarray}\label{eq:Disc}
    \left|\sqrt{2}|A_{LD}|\sin\Delta\delta\right| = \big|A(K_L)_\text{absorptive}\big| = \left|\frac{1}{2i}{\rm Disc}(K_L\to\gamma\gamma\to\mu^+\mu^-)\right|\, .
\end{eqnarray}
This discontinutiy can be computed in a straightforward way using Cutkosky rules, which entail considering the cut intersecting both photon propagators. In this way the left part of the cut diagram is directly related to the measured rate of $K_L\to\gamma\gamma$, and the right $\gamma\gamma\to\mu^+\mu^-$ part of the diagram requires QED only, allowing to extract the magnitude of the discontinuity completely model independently. 
This leads to
\begin{align}
\cos^2\varphi_0 &= \frac{\big[A(K_L)_\text{absorptive}\big]^2}{|A(K_L)|^2}= C_{\mathrm{QED}}^2 \frac{\Gamma (K_L\rightarrow\gamma\gamma)}{\Gamma(K_L\rightarrow \mu^+\mu^-)}\,,
\end{align}
\emph{i.e.},
\begin{eqnarray}\label{eq:master}
 \cos^2\varphi_0 \,   {\cal B}(K_L\to\mu^+\mu^-)\,  = \, C_\text{QED}^2 \, {\cal B}(K_L\to\gamma\gamma)\,,
\end{eqnarray}
where $C_{\mathrm{QED}}$ describes the $\gamma\gamma\rightarrow \mu^+\mu^-$ transition and is given as~\cite{Martin:1970ai}
\begin{equation}
    C_\text{QED} = \frac{\alpha_{em} m_\mu}{\sqrt{2\beta_\mu}m_K}\,\log\left(\frac{1-\beta_\mu}{1+\beta_\mu}\right) + {\cal O}(\alpha_{em}^2)\, .
\end{equation}
Equation~(\ref{eq:master}) demonstrates that $\cos^2\varphi_0$ is simply the proportionality factor parameterizing to what extent is the $K_L\to\mu^+\mu^-$ rate saturated by the absorptive contribution from the intermediate $\gamma\gamma$ state.
Using the measured ratio of rates $R_{K_L}$, see Eq.~\eqref{eq:RKL},
we have a clean SM prediction for the phase $\varphi_0$, up to a four-fold discrete ambiguity,
\begin{eqnarray}\label{eq:cleanPred}
    \cos^2\varphi_0 \, &=& \,  \frac{C_{\mathrm{QED}}^2}{R_{K_L}^{\mathrm{exp}}} \, .
\end{eqnarray}
This SM prediction is dependent only on
\begin{enumerate}
\item The measurement of $R_{K_L}$, currently with an uncertainty of ${\cal O}(2\%)$; 
\item A QED calculation, here taken up to relative corrections of ${\cal O}(\alpha_{em})$;
\item The assumption that other intermediate on-shell contributions ($3\pi$, $\pi\pi\gamma$) are negligible~\cite{Martin:1970ai}.
\end{enumerate}

Inserting
\begin{eqnarray}
    C_{\rm QED}^2 = 1.195 \times 10^{-5} \big[1+{\cal O}(\alpha_{em})\big], \label{eq:CQED}
\end{eqnarray}
and
\begin{eqnarray}
	R_{K_L}^{\rm exp}  = (1.250 \pm 0.022)\times 10^{-5}\, ,
\end{eqnarray}
which we obtain, not taking into account any correlations, from~\cite{Workman:2022ynf}
\begin{align}
\label{eq:expKLmumu} 
\mathcal{B}(K_L\rightarrow \mu^+\mu^- ) &= (6.84\pm 0.11)\times 10^{-9}\,, \\
\mathcal{B}(K_L\rightarrow \gamma\gamma) &= (5.47\pm 0.04)\times 10^{-4}\,, 
\end{align}
using Gaussian error propagation,
we arrive at
$\cos^2\varphi_0 = 0.96 \pm 0.02$ ,
where the quoted error reflects only the experimental error on $R_{K_L}^{\rm exp}$. 

There are two sources of theoretical errors. One of the them is the higher order QED calculation, resulting in an error of order $\alpha_{em} \sim 1\%$. This error is reducible, that is, if needed the calculation of the higher-order corrections can be done. The other source of uncertainty are the intermediate states that we neglected, such as $3\pi$ and $\pi\pi\gamma$~\cite{Martin:1970ai}. These contributions are estimated to be at most $1\%$ of the two-photon state that we considered. Each of the two  effects results in about $1\%$ error, and thus we conservatively add them linearly resulting in a total theory error of $0.02$ to arrive at our final estimate,
\begin{equation}
    \cos^2\varphi_0 = 0.96 \pm 0.02_{\mathrm{exp}}\pm 0.02_{\mathrm{th}}\, .
\end{equation}
Combining the experimental and theoretical errors in quadrature we then have
\begin{equation}
    \cos^2\varphi_0 = 0.96 \pm 0.03. \label{eq:combined-errors}
\end{equation}
Note that the error quoted in Eq.~(\ref{eq:combined-errors}) is therefore to be interpreted as an estimate of the total uncertainty, rather than as a statistical error only.

Given the value of $\cos^2\varphi_0$ there are four possible values for the phase shift, $\varphi_0$. Two of them correspond to overall constructive interference in the time-dependent rate and two correspond to destructive interference, depending on the sign of $\cos\varphi_0$. We plot the time-dependent rate for the four possibilities in Fig.~\ref{fig:cos4timeDep4} for both a $K^0$ or a $\overline{K}^0$ beam. Note that, for a $K^0$  beam, a positive (negative) $\cos\varphi_0$ results in constructive (destructive) interference.
As pointed out in Ref.~\cite{DAmbrosio:2017klp}, the situation is reversed for a $\overline{K}^0$ beam where 
a negative (positive) $\cos\varphi_0$ results in constructive (destructive) interference.
\begin{figure}[t!]
\centering
\subfigure[]{\includegraphics[width=0.3\textwidth]{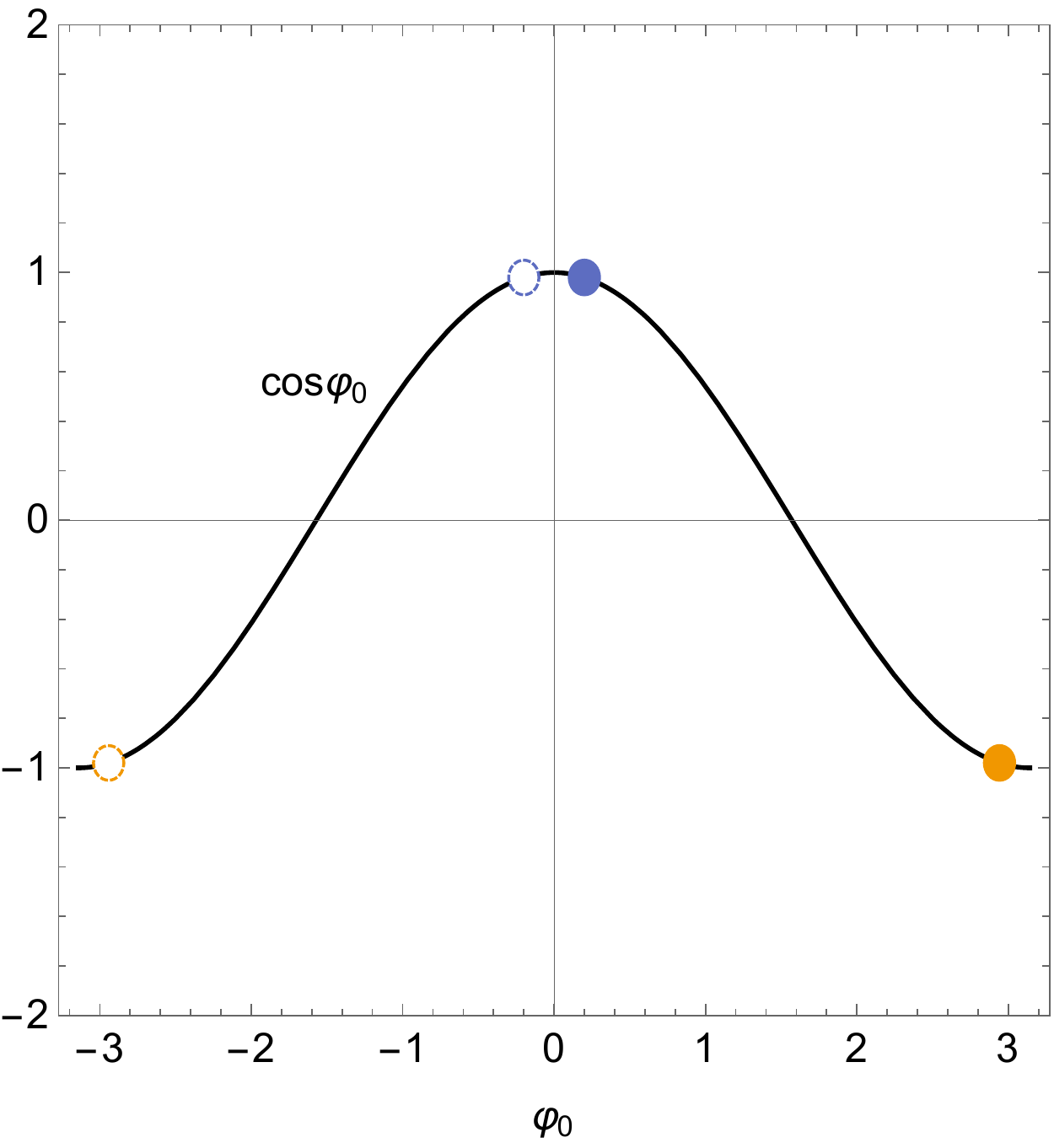}}
\subfigure[]{\includegraphics[width=0.32\textwidth]{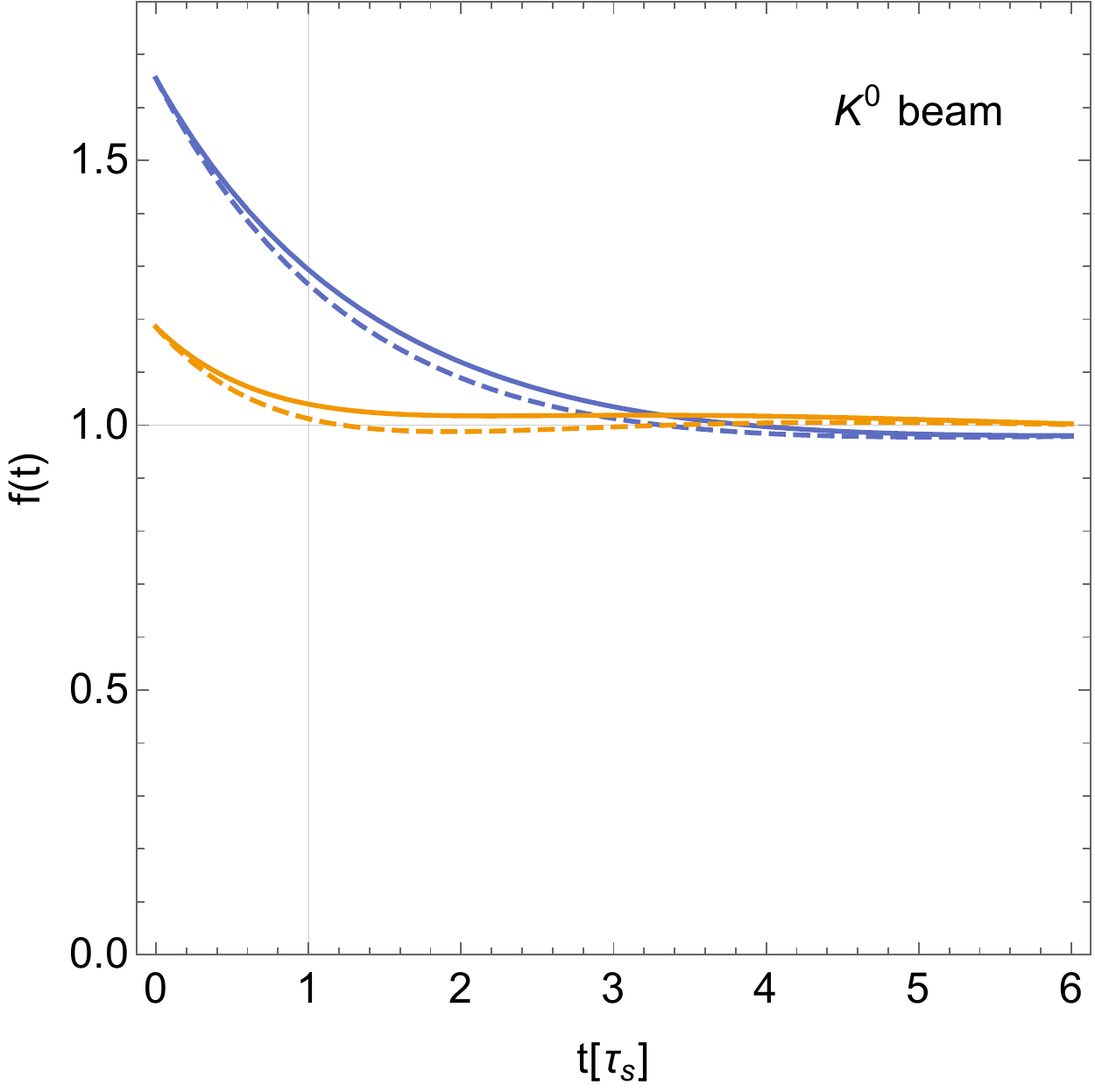}}
\subfigure[]{\includegraphics[width=0.32\textwidth]{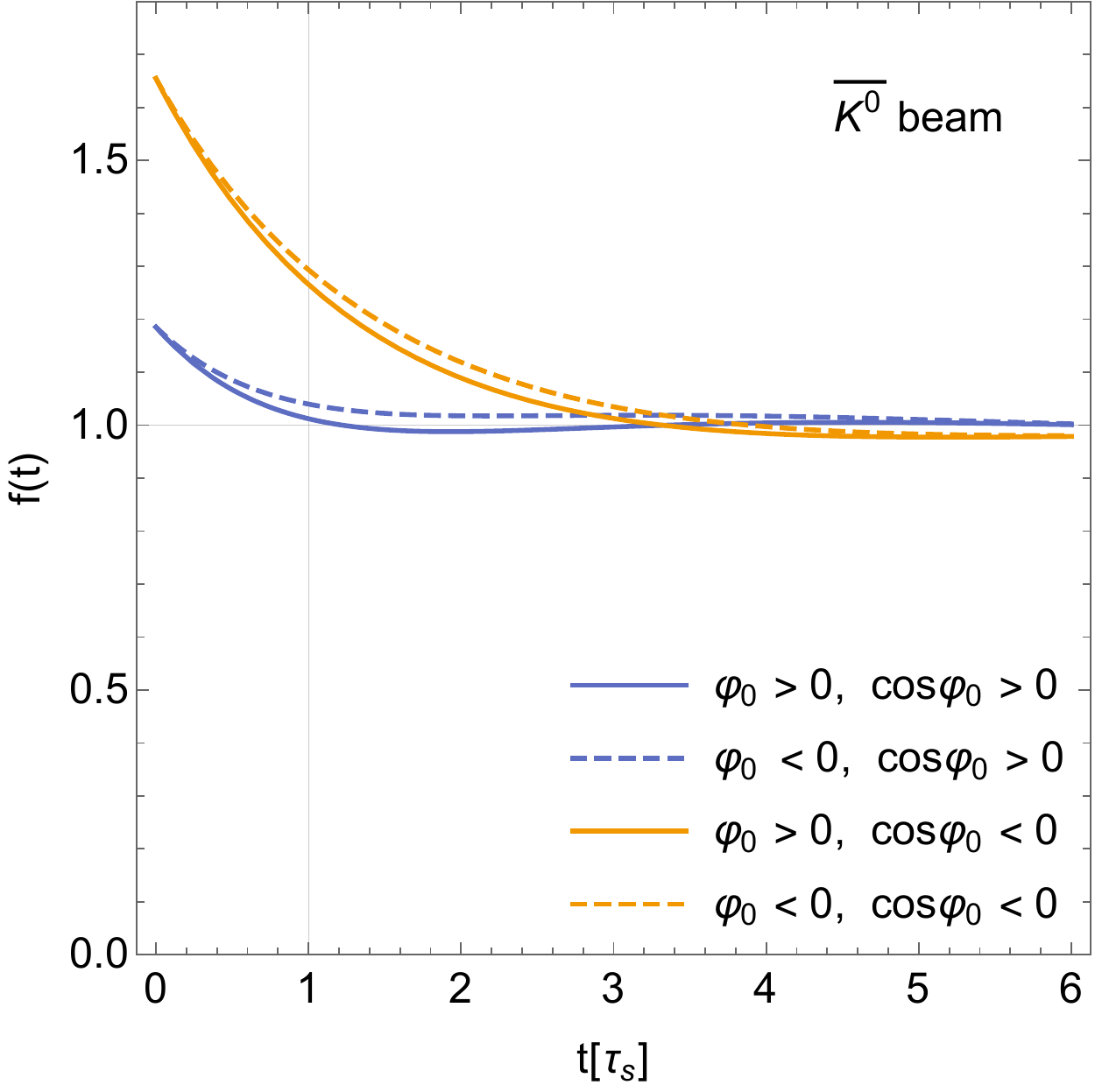}}
\caption{(a) Illustration of the four-fold ambiguity arising from $\cos^2\varphi_0 = 0.96$. (b) The time dependence of the  $K^0(t)\to\mu^+\mu^-$ rate, $f(t)$, as defined in Eq.~(\ref{eq:timeDep}), for the four values of $\varphi_0$. We have used a normalization in which $C_L$ of Eq.~(\ref{eq:timeDep}) is set to unity. (c) The same for an initial pure $\overline K^0$ beam.} 
\label{fig:cos4timeDep4}
\end{figure}
%

%%%%%%%%%%%%%%%%%%%%%%%%%%%%%%
\section{Determination of \texorpdfstring{\boldmath{$\cos\varphi_0$}}{cos phi0}:
model dependent
\label{sec:largeNC}}
%%%%%%%%%%%%%%%%%%%%%%%%%%%%%%
%

The four-fold ambiguity cannot be resolved in a model-independent way. Thus, in this section we use input from theory in order to try to reduce it. We first consider the sign of $\cos\varphi_0$. We discuss below how chiral perturbation theory (ChPT) and lattice QCD can help in this regard.

\subsection{Chiral perturbation theory}
Within chiral perturbation theory and using the large-$N_C$ limit, it has been shown that the sign of the absorptive amplitude relative to the short-distance contribution, which determines the sign of $\cos\varphi_0$, can be determined~\cite{GomezDumm:1998gw,Isidori:2003ts}.

In the following, we relate our notation for the long-distance contribution to that appearing in the literature in order to apply existing results, primarily of Refs.~\cite{Isidori:2003ts,GomezDumm:1998gw,Knecht:1999gb,DAmbrosio:1997eof}. 
We rewrite the long-distance amplitude for $K_L\to\mu^+\mu^-$ as
\begin{align}\label{eq:ChadDef}
    \sqrt{2}|A_{LD}|e^{i\Delta\delta} \, &= \, \big[A_{LD}\big]_{\rm dispersive} + i\big[A_{LD}\big]_{\rm absorptive}\, \nonumber \\
    &\equiv \, C_{\rm had.} \Big[\big(A_{LD}^{\rm local} + {\rm Re}A_{LD}^{\gamma\gamma}\big) + i\,{\rm Im}A_{LD}^{\gamma\gamma}\Big]\, ,
\end{align}
where $C_{\rm had.}$ encodes the hadronic behavior of the effective $K_L\gamma\gamma$ vertex.
Without loss of generality, we take $C_{\rm had.}$ to be real, consistent with Eq.~\eqref{eq:AKL-absorptive}.
The dispersive part is split in two parts to be consistent with the literature, 
\begin{align}
\big[A_{LD}\big]_{\rm dispersive} &= C_{\rm had.} \left( A_{LD}^{\rm local} + {\rm Re}A_{LD}^{\gamma\gamma} \right)\,,\\
\big[A_{LD}\big]_{\rm absorptive} &= C_{\rm had.} \,  {\rm Im}A_{LD}^{\gamma\gamma}   \,,
\end{align}
where $A_{LD}^{\rm local}$ denotes the local counterterm, which is real \cite{Isidori:2003ts}. 
Using the fact that the decay $K_L\to\gamma\gamma$ has the same hadronic behavior, we define
\begin{equation}
    A(K_L\to\gamma\gamma) \equiv C_{\rm had.}\,A_{\gamma\gamma}\, .
\end{equation}
Note that $A_{\gamma\gamma} \neq A^{\gamma\gamma}_{LD}$. The former is part of the $K_L\rightarrow \gamma\gamma$ amplitude, while the latter is the two-photon contribution to $K_L\rightarrow \mu^+\mu^-$.  
Then we have 
\begin{eqnarray}
\frac{\Gamma(K_L\to\mu^+\mu^-)}{\Gamma(K_L\to\gamma\gamma)} \, &=& \, 
2\beta_{\mu}
\left|\frac{\sqrt{2} |A_{LD}|e^{i\Delta\delta} + \sqrt{2}|A_{SD}|\cos\theta_{SD}}{A(K_L\to\gamma\gamma)}\right|^2 \\ \nonumber
&=& 2\beta_{\mu} \frac{\left(A_{LD}^{\rm local}+{\rm Re}A_{LD}^{\gamma\gamma}+ \sqrt{2}|A_{SD}|\cos\theta_{SD}/C_{\rm had.}\right)^2 + ({\rm Im}A_{LD}^{\gamma\gamma})^2 }{|A_{\gamma\gamma}|^2}\, .
\end{eqnarray}
Note that $C_{\rm had.}$ now appears as a factor accompanying the short-distance contribution.
We can now easily relate to the notations of Ref.~\cite{Isidori:2003ts}, with (the superscript ``IU" denotes the initials of the authors of Ref.~\cite{Isidori:2003ts})
\begin{align}
    \frac{A_{LD}^{\rm local}}{|A_{\gamma\gamma}|}  &=  
    \frac{\alpha_{em} m_{\mu}}{\pi m_K}
    \big[\chi_{\gamma\gamma}(\mu)\big]^{\rm IU}\,, \\
    %%%%
    \frac{{\rm Re}A_{LD}^{\gamma\gamma}}{|A_{\gamma\gamma}|} &= 
    \frac{\alpha_{em} m_{\mu}}{\pi m_K}
    \left[{\rm Re}C_{\gamma\gamma} -\frac{5}{2} + \frac{3}{2}\log\left(\frac{m_\mu^2}{\mu^2}\right)\right]^{\rm IU}\, , \\ 
    %%%%
    \frac{\sqrt{2}|A_{SD}|\cos\theta_{SD}}{C_{\rm had.}|A_{\gamma\gamma}|} \, &=\, 
    \frac{\alpha_{em} m_{\mu}}{\pi m_K}
    \big[\chi_{\rm short}\big]^{\rm IU}\, , \\
    %%%%
    \frac{{\rm Im}A_{LD}^{\gamma\gamma}}{|A_{\gamma\gamma}|} &= 
    \frac{\alpha_{em} m_{\mu}}{\pi m_K}
    \big[{\rm Im}C_{\gamma\gamma}\big]^{\rm IU} = \frac{1}{\sqrt{2\beta_{\mu}}} C_{\rm QED} \,.
\end{align}

The assumptions and findings of the literature, as conveyed in Refs.~\cite{Isidori:2003ts,GomezDumm:1998gw}, can now be summarized as the following:
\begin{enumerate}
    \item Using results in the large-$N_C$ limit, Refs.~\cite{Isidori:2003ts,GomezDumm:1998gw} find destructive interference between the short-distance and the local long-distance contributions,
    \begin{equation}\label{eq:LDSDint}
        \frac{\cos\theta_{SD}}{C_{\rm had.}A_{LD}^{\rm local}} \, < \, 0\, .
    \end{equation}

    \item Ref.~\cite{Isidori:2003ts} uses phenomenological analyses of the form factor in $K_L\to\gamma e^+ e^-$, $K_L\to\gamma\mu^+\mu^-$ and $K_L\to e^+e^-\mu^+\mu^-$ from data, together with theory considerations, to estimate the local counter term. Using up-to-date inputs, we update their estimation (see App.~\ref{app:chi}) and find
    \begin{equation}\label{eq:chigamgamIU}
         \big[\chi_{\gamma\gamma}(m_\rho)\big]^{\rm IU} \, = \, {\frac{\pi m_K}{ \alpha_{em} m_{\mu}}}  \frac{A_{LD}^{\rm local}}{|A_{\gamma\gamma}|} = (6.10 \pm 1.01)\, > \, 0 \, ,
    \end{equation}
    where we have set $\mu^2=m_\rho^2$ here and in the following whenever we make use of specific numerical estimates.
    Hence, using Eq.~(\ref{eq:LDSDint}),
    \begin{equation} \label{eq:sgnChad}
        {\rm sgn}\big[C_{\rm had.}\big] \, = \, -{\rm sgn}\big[\cos\theta_{SD}\big]\, .
    \end{equation}

    \item This, in turn, determines the sign of the absorptive long-distance amplitude relative to the short-distance contribution,
    \begin{eqnarray}
        {\rm sgn}\left(\big[A_{LD}\big]_{\rm absorptive}\right) \, &=& \, {\rm sgn}\big[\cos\theta_{SD}\big] \, ,
    \end{eqnarray}
    where we used the fact that ${\rm sgn}\left[{\rm Im}A_{LD}^{\gamma\gamma}\right] = {\rm sgn}\left[C_{\rm QED}\right] = -1$.
\end{enumerate}
We conclude that within a model for the short-distance contribution, and adopting the assumptions in the literature regarding the long-distance physics, \ie, the large-$N_C$ limit, the sign of $\cos\varphi_0$ is determined,
\begin{eqnarray}\label{eq:sgncosphi0}
    {\rm sgn}\Big[\cos\varphi_0\Big] \, &=& \, {\rm sgn}\left[\frac{{\rm Re}\big[A(K_S)^*A(K_L)]}{|A(K_S)A(K_L)|} \right] \\ \nonumber
    &=&\,  {\rm sgn}\left[\frac{\big[A(K_L)\big]_{\rm absorptive}\, {\rm sgn}[\sin\theta_{SD}]}{|A(K_L)|}\right] \\ \nonumber
    &=& \, {\rm sgn}\Big[\tan\theta_{SD}\Big]\, .
\end{eqnarray}

%%%%%%%%%%%
\subsection{Detailed assumptions within ChPT}
\label{sec:ChPT}
The considerations leading to the assumption of destructive interference between the short-distance and the local long-distance contributions, as conveyed in Eq.~(\ref{eq:LDSDint}), involve some details of the structure of the $K_L\to\gamma\gamma$ amplitude within the ChPT. 
According to Ref. \cite{Cirigliano:2011ny},
the on-shell tensor amplitude for $K_L \to \gamma \gamma$ starts from at $\mathcal{O}(p^6)$ in the ChPT as
\beq
-iA(K_L \to \gamma \gamma)
=  \varepsilon^{\mu \nu \rho \sigma}\epsilon_{1\mu} (q_1) \epsilon_{2\nu} (q_2) q_{1\rho}q_{2\sigma} \, c^{(6)}(0,0)\,,
\eeq
where $c^{(6)}(q_1^2, q_2^2)$ is an amplitude for $K_L \to \gamma^\ast (q_1) \gamma^\ast (q_2)$.
We factorize the $\pi^0$ exchange contribution in $c^{(6)}(0,0)$ and define a dimensionless reduced amplitude as $c_{\rm red}^{(6)}$, 
\beq
c^{(6)}(0,0)\, =\,  - \frac{2}{\pi} \alpha_{em} F_0 \left( G_8 - G_{27}\right) \frac{1}{1-r_\pi^2} c_{\rm red}^{(6)}\,,
\eeq
%with $r_P =  m_P /m_K$, 
and $c_{\rm red}^{(6)}$ can be expanded in the ChPT as
\begin{align}
\begin{aligned}
c_{\rm red}^{(6)}\, = \, & 1 +\frac{1-r_\pi^2 }{3\left(1-r_\eta^2\right)}
\left[ \left(1+\xi\right) c_\theta + 2 \sqrt{2} \, \hat\rho \, s_\theta\right]\left(\frac{F_\pi}{F_{\eta_8}}
c_\theta - 2 \sqrt{2} \frac{F_\pi}{F_{\eta_1}}s_\theta \right) \\
& - \frac{1-r_\pi^2 }{3\left(1-r_{\eta^\prime}^2\right)}
\left[ 2 \sqrt{2} \, \hat\rho\, c_\theta - \left(1+\xi\right) s_\theta  \right]\left(\frac{F_\pi}{F_{\eta_8}}
s_\theta + 2 \sqrt{2} \frac{F_\pi}{F_{\eta_1}}c_\theta \right)\,,
\end{aligned}
\end{align}
with $\sqrt{2} F_0 = \sqrt{2} F_{\pi} = f_{\pi}=(130.2\pm 0.8)$ MeV \cite{FlavourLatticeAveragingGroupFLAG:2021npn}, 
 $F_{\eta_8} = (1.27 \pm 0.02)\, F_{\pi}$,
  $F_{\eta_1} = (1.14 \pm 0.05)\, F_{\pi}$ \cite{Escribano:2015yup,Gan:2020aco}, and $r_P\equiv m_P/m_K$.
Combining the above formulae gives 
\beq
- iA(K_L \to \gamma \gamma)
=  - \frac{\sqrt{2}\alpha_{em} f_\pi \left( G_8 - G_{27}\right) }{\pi \left(1-r_\pi^2\right)}c_{\rm red}^{(6)} \varepsilon^{\mu \nu \rho \sigma}\epsilon_{1\mu} (q_1) \epsilon_{2\nu} (q_2) q_{1\rho}q_{2\sigma}  \,.
\eeq
Here, the point is that $\mathcal{O}(p^4)$ contributions  vanish within the ChPT,
which are proportional to $c^{(4)}_{\rm red}$ and 
\beq
 c_{\rm red}^{(4)} \, = \,
1 +\frac{1-r_\pi^2 }{3\left(1-r_{\eta_8}^2\right)} 
\, = \,  \frac{4 - 3 r_{\eta_8}^2 - r_{\pi}^2 }{3\left(1-r_{\eta_8}^2\right)} \,= \, 0\,,
\eeq
where the Gell-Mann--Okubo mass formula, $
4 m_K^2 = 3 m_{\eta_8}^2 + m_\pi^2$, 
is used. 
Therefore, 
$c_{\rm red}^{(6)}$ amplitudes correspond to the violation of the  Gell-Mann--Okubo formula implying that 
the sign of $c_{\rm red}^{(6)}$ is sensitive to the $\eta$--$\eta^\prime$ mixing angle $\theta$ in the octet--singlet basis, 
the $SU(3)_F$ breaking $
\xi$ \cite{Donoghue:1986ti,He:2002as}, 
the nonet symmetry breaking $\hat{\rho}$ \cite{Donoghue:1983hi,Cheng:1990mw,DAmbrosio:1997hlp}, 
and their higher-order corrections. 
$(1+\xi)$ is proportional to the $K_L \to \eta_8$ form factor, while $\hat{\rho}$ is proportional to the $K_L \to \eta_1$ one.
However, 
by considering the typical parameter regions;
$\theta \approx - 20^\circ$, $\xi \sim 0.0\text{--}0.2$, and $\hat{\rho}\approx 0.8$~\cite{Cirigliano:2011ny}, 
one can predict $\text{sgn}[c_{\rm red}^{(6)}]>0$,
which leads to
\begin{align}
    {\rm sgn}\Big[A(K_L\to\gamma \gamma)\Big] 
=  {\rm sgn}\Big[A(K_L\to\pi^0 \to \gamma \gamma)\Big] \,.
\end{align}
Combining this relation with 
$\text{sgn}[G_8-G_{27}]$ which can be extracted 
from the $\Delta S=1$ effective Lagrangian in the large-$N_C$ limit  \cite{Pich:1995qp,Isidori:2003ts,Gerard:2005yk}, 
%fits to $K\rightarrow \pi\pi$ and lattice QCD data,
 $\text{sgn}[C_{\rm had.}] > 0$ can be predicted, see Eq.~(\ref{eq:sgnChad}), where, in the SM we have $\cos\theta_{SD}^{\mathrm{SM}} <0$.

\subsection{Lattice QCD}

In the last decade, lattice QCD made paramount progress in the treatment of $K\rightarrow \pi\pi$~\cite{RBC:2020kdj, RBC:2015gro, Blum:2015ywa, Blum:2011ng}. Moreover, recently, lattice QCD made advances in the calculation of $K_L\rightarrow \mu^+\mu^-$ and $K_L\rightarrow \gamma\gamma$~\cite{Christ:2020bzb, Zhao:2022pbs, Christ:2022rho}.
ChPT parameters like $G_8$ can now also be extracted from fits to lattice results, as shown in Ref.~\cite{Pich:2021yll}. 

While it seems to us that the data is available to extract the sign, we were unable to find it from the available publications. It would be interesting to use the available lattice data to obtain it. Such an extraction would be interesting to confront the ChPT results.

%%%%
\subsection{SM prediction for the short-distance physics}
Within the SM, the short distance contribution arises from the following effective Hamiltonian~\cite{Buchalla:1995vs}
\beq\label{eq:SMeffH}
{\cal H}_{\rm eff} = -\frac{G_F}{\sqrt{2}}\frac{\alpha_{em}}{2\pi\sin^2\theta_W} \left[ V_{ts}^*V_{td}Y(x_t)+V_{cs}^*V_{cd} Y_{NL}\right] \left[(\bar s d)_{V-A}(\bar \mu \mu)_{V-A}\right] + \text{h.c.}\, .
\eeq
We can then write (in the basis where $\delta_{SD}=0$)
\begin{eqnarray}\label{eq:ASD}
    A(K_L)_{SD}^{\rm SM}\, &=& \, \big(\sqrt{2}|A_{SD}|\cos\theta_{SD}\big)^\text{SM} \\ \nonumber
    \, &=& \,    \frac{\sqrt{2} G_F \alpha_{em}}{\pi\sin^2\theta_W}\big|V_{ts}^*V_{td}Y(x_t)+V_{cs}^*V_{cd}Y_{NL}\big| f_K m_\mu m_K \,\cos\theta_{SD}^{\rm SM}\, ,
\end{eqnarray}
where we identify
\begin{equation}
    \theta_{SD}^\text{SM}\, = \, \arg\left(-\frac{V_{ts}^*V_{td} + V_{cs}^*V_{cd}Y_{NL}/Y(x_t)}{V_{cs}^*V_{cd} }\right)\, .
\end{equation}
We therefore find,
\begin{eqnarray}
    \tan\theta_{SD}^{\rm SM} \, &=& \, 
    -\frac{\eta}{(1-\rho)+\frac{1}{A^2\lambda^4}\frac{Y_{NL}}{Y(x_t)}} + {\cal O}(\lambda^6) \, < \, 0 \, .
\end{eqnarray}

Hence, from Eq.~(\ref{eq:sgncosphi0}), within the SM and under the aforementioned model-dependent assumptions, we have
\begin{eqnarray}
    \big[\cos\varphi_0\big]^{\rm SM}_{\text{large}\, N_C} \, < \, 0 \, .
\end{eqnarray}
Together with the result of section~\ref{sec:model-independent}, $\cos^2\varphi_0 = 0.96 \pm 0.03$, we have
\begin{eqnarray}
\big[\cos\varphi_0\big]^{\rm SM}_{\text{large}\, N_C} \, = \, -0.98 \pm 0.02 \, .
\end{eqnarray}
Note that the error combines both a statistical error from experiment as well as a theory component, \emph{i.e.}~is to be interpreted as an estimate of the total uncertainty.

%%%%%%%%%%%%%%%%%%%%%%
\section{Going beyond the two-fold ambiguity   \label{sec:determination}}
%%%%%%%%%%%%%%%%%%%%%%

In order to determine the sign of $\sin\varphi_0$ and get rid of the remaining ambiguity, we would need to determine the signs and magnitudes of the competing short-distance and long-distance dispersive contributions,
\begin{align}
    \sin\varphi_0  &= \frac{{\rm Im}\big[A(K_S)^* A(K_L)\big]}{|A(K_S)A(K_L)|}\nn\\ 
    &= -\frac{\left(\big[A_{LD}\big]_{\rm dispersive}+\sqrt{2}|A_{SD}|\cos\theta_{SD}]\right)\mathrm{sgn}[\sin(\theta_{SD})]}{|A(K_L)|}\,.\label{eq:sinGeneral}
\end{align}
We recall that $\big[A_{LD}\big]_{\rm dispersive}$ can be written as  
\begin{align}
\big[A_{LD}\big]_{\rm dispersive} &= \sqrt{2} \vert A_{LD}\vert \cos\Delta\delta = C_{\rm had.} \left( A_{LD}^{\rm local} + {\rm Re}A_{LD}^{\gamma\gamma} \right) \,,
\end{align}
see Eq.~(\ref{eq:ChadDef}). 
Existing semi-phenomenological theory estimations of $\big[A_{LD}\big]_{\rm dispersive}$ come with large theory uncertainties.
Using the estimate of Ref.~\cite{Isidori:2003ts} as in Eq.~(\ref{eq:chigamgamIU}), updated with existing data (see App.~\ref{app:chi}), we have for the long-distance contribution, 
\begin{eqnarray} \label{eq:LDdispNum}
    \frac{\big[A_{LD}\big]_{\rm dispersive}}{|A(K_L\to\mu^+\mu^-)|} \, &=& \, \frac{C_{\rm had.}}{|A(K_L\to\mu^+\mu^-)|}\left(A_{LD}^{\rm local} + {\rm Re}A_{LD}^{\gamma\gamma}\right) \\
    && \hspace*{-34mm}\, = \frac{C_{\rm had.}|A_{\gamma\gamma}|}{|A(K_L\to\mu^+\mu^-)|} \frac{\alpha_{em}m_\mu}{\pi m_K} \left(\big[\chi_{\gamma\gamma}(m_\rho)\big]^{\rm IU} + \left[{\rm Re}C_{\gamma\gamma}-\frac{5}{2}+ \frac{3}{2}\log\left(\frac{m_\mu^2}{m_\rho^2}\right)\right]^{\rm IU}\right)\nonumber \\ 
    && \hspace*{-34mm} =\, \text{sgn}\left[ C_{\rm had.}\right] \sqrt{\frac{2\beta_\mu}{R_{K_L}}} \frac{\alpha_{em}m_\mu}{\pi m_K}\left[ (6.10 \pm 1.01) -5.14\right]
    \in\,\big[ -0.009,\, 0.37\big] \,,\nonumber 
\end{eqnarray}
where $\text{sgn}[C_{\rm had.}]>0$ derived in the previous section is used.

For the short-distance SM contribution, we have (see Eq.~\eqref{eq:ASD}) 
\begin{eqnarray}\label{eq:SDNum}
    \frac{A(K_L)_{SD}^{\rm SM}}{|A(K_L\to\mu^+\mu^-)|}\, &=& \, \frac{\sqrt{2} G_F \alpha_{em}(m_Z)}{\pi\sin^2\theta_W}\frac{\big|V_{ts}^*V_{td}Y(x_t)+V_{cs}^*V_{cd}Y_{NL}\big| f_K m_\mu m_K \,\cos\theta_{SD}^{\rm SM}}{|A(K_L\to\mu^+\mu^-)|}\\ \nonumber
    &=& \,-0.331 \pm 0.008\, ,
\end{eqnarray}
where we use the following inputs,
\begin{eqnarray}
    &\,& Y(x_t)=0.931 \pm 0.005~\text{\cite{Brod:2022khx}} \, , \qquad\qquad  Y_{\rm NL} = (2.95\pm 0.46)\times  10^{-4}~\text{\cite{Gorbahn:2006bm}}\, ,\\ \nonumber
    &\,& A = 0.790^{+0.017}_{-0.012}\, ,\qquad\qquad\qquad\quad\qquad \lambda = 0.22650\pm 0.00048 \, , \\ \nonumber
    &\,& \bar\rho = 0.141^{+0.016}_{-0.017} \,, \qquad\qquad\qquad\quad\,\,\,\,\qquad \bar\eta =0.357\pm 0.011 \, ,\\ \nonumber
    &\,& m_K = 497.61\, {\rm MeV}\, , \qquad\qquad \qquad\,\,\,\quad m_\mu = 105.658\, {\rm MeV}\, ,\\ \nonumber
    &\,&G_F = 1.166378\times 10^{-5} \, {\rm GeV}^{-2}\, ,\quad \quad \,\,\, f_K = 155.7\, {\rm MeV}\, , \\ \nonumber 
    &\,&\alpha_{em} = 1/137\, , \qquad \qquad \qquad \qquad \qquad \alpha_{em}(m_Z) =1/129 \, , \\ \nonumber 
    &\,&\sin^2\theta_W = 0.23\,, \qquad\qquad \qquad \,\,\quad \qquad m_{\rho} = 775.26\,\text{MeV}\, .
\end{eqnarray}
In particular, the CKM input results in $\cos\theta_{SD}^{\mathrm{SM}} = -0.94$.

Hence, the large theory uncertainty on the dispersive long-distance contribution, as reflected in the range given in Eq.~(\ref{eq:LDdispNum}), does not allow to determine if it is larger or smaller than the short-distance SM contribution, Eq.~(\ref{eq:SDNum}).
We conclude that with current knowledge on the dispersive long-distance contribution the sign of $\sin\varphi_0$ cannot be determined.

Therefore, we have, using Gaussian error propagation,
\begin{eqnarray}
    \big[\cos\varphi_0\big]^{\rm SM}_{\text{large}\, N_C} \, = \, -0.98 \pm 0.02 \, , \qquad \big[\sin\varphi_0\big]^{\rm SM}_{\text{large}\, N_C,\,\text{theory}} \, = \, \pm \big(0.21 \pm 0.07\big)\, . \label{eq:sinphi-result}
\end{eqnarray}
We note that due to the nature of the theoretical error, the error should not be interpreted as a statistical error but rather as an estimate of the uncertainty.

%%%%%%%%%%%%%%
\section{Discussion and Conclusions \label{sec:conclusion}}

In this work we have related the phase shift, $\varphi_0$, appearing in the time-dependent decay rate of a neutral kaon to a dimuon pair to the ratio of integrated rates of $ K_L \to \mu^+ \mu^-$ and $K_L \to \gamma \gamma$. This relation holds to an excellent approximation under the well-motivated assumption that the two-photon intermediate state dominates the absorptive contribution, and within any model in which the short-distance leptonic current is axial or vectorial, as in the SM. The only input required other than the ratio of integrated rates is a coefficient calculated within QED. We find that $\cos^2 \varphi_0$ is precisely predicted model independently, and given by
\begin{equation}
    \cos^2\varphi_0 \, = \, 0.96 \pm 0.02_{\mathrm{exp}} \pm 0.02_{\mathrm{th}}\,.
\end{equation}
The experimental error comes from the error of $R_{K_L}$ and the theory error is our estimate of the size of higher order QED corrections and the contribution from other intermediate states beside the di-photon state.
The result leaves a four-fold ambiguity in $\varphi_0$. 

The phase shift, $\varphi_0$, is also of experimental significance since it controls the integrated number of interference events. For a $K^0$ beam a positive value of $\cos\varphi_0$ is preferred as it enhances the interference and improves the feasibility of extracting clean short-distance information from the interference term. For a $\overline{K}^0$ beam the situation is reversed, and a negative value of $\cos\varphi_0$ is preferred.
Thus, for experiments employing a proton beam on target, where the number of $K^0$ particles is expected to exceed that of $\overline{K}^0$ particles, a positive $\cos\varphi_0$ would be preferred.

We were unable to  determine the sign of $\cos \varphi_0$ completely model independently. Thus, with the use of several assumptions, that is, within the framework of ChPT and using a typical parameter region (motivated by the large-$N_C$ limit, as detailed in Sec.~\ref{sec:ChPT}), the sign of $\cos\varphi_0$, relative to the short-distance contribution, can be predicted. We find that within this framework, and assuming that the short-distance contribution is SM-like, there is a theory preference towards  $\cos\varphi_0 <0$. 
New Physics can potentially yield a different sign for $\cos \varphi_0$. A measurement of the angle $\varphi_0$ is therefore a test of the validity of several assumptions, pertaining to both the short-distance physics and the ChPT description of the dispersive long-distance physics.

Given the assumptions that were made to arrive to the conclusion about the sign of $\cos\varphi_0$, and the fact that the prediction can be modified for models beyond the SM, we conclude that
neither solution is unequivocally theoretically favored.
That is,
we cannot conclude that we know the sign of $\cos\varphi_0$ to high confidence. Thus, when planing to perform the experiment we encourage the experimental collaborations to consider both possible signs for $\cos\varphi_0$ for purposes of sensitivity estimations.

We therefore conclude by emphasizing that the time dependence of the kaon decay rate to two muons provides two very clean SM predictions:
\begin{enumerate}
\item  
The coefficient of the interference term allows the extraction of the theoretically clean decay rate ${\cal B}(K_S \to \mu^+ \mu^-)_{\ell=0}$. In the SM, this observable is proportional to the CKM combination $|V_{ts} V_{td} \sin (\beta + \beta_s)|$. 
\item
Although it includes long-distance as well as short-distance physics, the phase shift in the interference term, $\varphi_0$, is predicted cleanly up to a four-fold ambiguity.
\end{enumerate}
Thus, an experiment that performs the time-dependence studies of the $K \to \mu^+ \mu^-$ decay rate  provides two independent tests of the SM from the same measurement. While the phase $\varphi_0$ is not determined model independently, it directly impacts the measurement of the $K\to \mu^+\mu^-$  time dependent rate as it affects the interference between the $K_S$ and $K_L$ amplitudes in the total rate. 
The phase shift $\varphi_0$ is therefore a quantity of critical importance in kaon physics, as a way to test the SM, and in extension, as a probe of new physics beyond the SM because of its sensitivity to short distance effects. 
%%%%%%%%%%%%%%%
%

\begin{acknowledgments}
We thank Joachim Brod, Hector Gisbert, Antonio Pich, and Emmanuel Stamou for useful discussions.
Y.G.\ is supported in part by the NSF grant PHY1316222.
S.S.~is supported by a Stephen Hawking Fellowship from UKRI under reference EP/T01623X/1 and the Lancaster-Manchester-Sheffield Consortium for Fundamental Physics, under STFC research grant ST/T001038/1.
The work of T.K.\ is supported by the Japan Society for the Promotion of Science (JSPS) Grant-in-Aid for Early-Career Scientists (Grant No.\,19K14706) and the JSPS Core-to-Core Program (Grant No.\,JPJSCCA20200002). 
For the purpose of open access, the authors have applied a Creative Commons Attribution (CC BY) licence to any Author Accepted Manuscript version arising.
This work uses existing data which is available at locations cited in the bibliography.
\end{acknowledgments}

%%%%%%%%%%%%%%%%%%%%%%
\appendix

\section{Update of  the theory estimate for 
\texorpdfstring{\boldmath{$\chi_{\gamma\gamma}(\mu)$}}{chi gamma gamma (mu)}
}
\label{app:chi}
In this section we update the theory estimate of Ref.~\cite{Isidori:2003ts}.
The experimental input that goes into this estimate comes from analyses of the form factor in $K_L\to\gamma e^+ e^-$, $K_L\to\gamma\mu^+\mu^-$ and $K_L\to e^+e^-\mu^+\mu^-$, and can be summarized by the parameter $\alpha_{\rm exp.}$, defined by
\begin{equation}
    \alpha_{\rm exp.} = -m_\rho^2 \frac{d}{dq^2}f(q^2,0)\biggr\rvert_{q^2=0}\, ,
\end{equation}
where $f(q_1^2, q_2^2)$ is the $K_L\rightarrow \gamma\gamma$ form factor.
The experimental value derived from $K_L\to\gamma e^+e^-$ has been updated by the KTeV collaboration after Ref.~\cite{Isidori:2003ts} was published. We take~\cite{KTeV:2007ksh}
\begin{equation}
	\alpha_{\rm exp.}|_{e e} = -1.73 \pm 0.05\, ,
\end{equation}
combined with~\cite{KTeV:2001sfq,KTeV:2002kut}
\begin{eqnarray}
	\alpha_{\rm exp.}|_{\mu\mu} &=& -1.54\pm 0.10\, , \\ \nonumber
	\alpha_{\rm exp.}|_{e e\mu\mu} &=& -1.59 \pm 0.37\, ,
\end{eqnarray}
and arrive at the weighted average
\begin{equation}
	\alpha_{\rm exp.} = -1.691 \pm 0.044 \, .
\end{equation}
Comparing with Ref.~\cite{Isidori:2003ts} (who quote $\alpha_{\rm exp.}=-1.611\pm 0.044$), the central value has gone up by $\sim 5\%$ while the relative error remains the same.

Inserting this into Eq.~(22) of Ref.~\cite{Isidori:2003ts}, we have
\begin{eqnarray}
	\chi_{\gamma\gamma}(m_\rho) = (6.10 \pm 0.16_{\rm exp.}) -\Delta_\Lambda\,,
\end{eqnarray}
(updated from $(5.83 \pm 0.15_{\rm exp.}) -\Delta_\Lambda$). 
Following Ref.~\cite{Isidori:2003ts} we take
\begin{equation}
	|\Delta_\Lambda| \leq 1.0 \, ,
\end{equation}
and reach the result
\begin{equation}\label{eq:updateIU}
	\Big[\chi_{\gamma\gamma}(m_\rho)\Big]_{\rm IU} = 6.10 \, \pm \, 0.16_{\rm exp.} \, \pm \, 1.0_{\rm th.} \, .
\end{equation}

\bibliographystyle{utphys28mod}

\bibliography{DGGS_phi0}

\end{document}